\input harvmac.tex





\let\d\partial
\let\vphi\varphi
\def\frac#1#2{{\textstyle{#1\over #2}}}

\def\eqalignD#1{
\vcenter{\openup1\jot\halign{
\hfil$\displaystyle{##}$~&
$\displaystyle{##}$\hfil~&
$\displaystyle{##}$\hfil\cr
#1}}
}
\def\eqalignT#1{
\vcenter{\openup1\jot\halign{
\hfil$\displaystyle{##}$~&
$\displaystyle{##}$\hfil~&
$\displaystyle{##}$\hfil~&
$\displaystyle{##}$\hfil\cr
#1}}
}

\global\newcount\tableno
\def\table#1#2{\global\advance\tableno by1
\par\nobreak
\centerline
{\vbox{\offinterlineskip
\hrule
\halign{#2}
\hrule}}
}

\def\text#1{\quad\hbox{#1}\quad}

\def\e{\epsilon}
\def\ta{\theta}
\def\a{\alpha}

\let\Rw\Rightarrow

\def\rw{\rightarrow}

\def\lra{\leftrightarrow}

\def\max{{\rm max}}

\newcount\eqnum
\eqnum=0
\def\eq{\eqno(\secsym\the\meqno)\global\advance\meqno by1}
\def\eqlabel#1{{\xdef#1{\secsym\the\meqno}}\eq }
\newwrite\refs
\def\startreferences{
\immediate\openout\refs=references
\immediate\write\refs{\baselineskip=14pt \parindent=16pt \parskip=2pt}
}
\startreferences
\refno=0
\def\aref#1{\global\advance\refno by1
\immediate\write\refs{\noexpand\item{\the\refno.}#1\hfil\par}}
\def\ref#1{\aref{#1}\the\refno}
\def\refname#1{\xdef#1{\the\refno}}

\newcount\exno
\exno=0
\def\Ex{\global\advance\exno by1{\noindent\sl Example \the\exno:
\nobreak\par\nobreak}}
\parskip=6pt
\overfullrule=0mm


\def\frac#1#2{{#1 \over #2}}

\let\ta=\theta
\let\d=\partial

\let\n=\noindent
\let\e=\epsilon

\def\rw{{\rightarrow}}

\newwrite\refs
\def\startreferences{
\immediate\openout\refs=references
\immediate\write\refs{\baselineskip=14pt \parindent=16pt \parskip=2pt}
}
\startreferences
\refno=0
\def\aref#1{\global\advance\refno by1
\immediate\write\refs{\noexpand\item{\the\refno.}#1\hfil\par}}
\def\ref#1{\aref{#1}\the\refno}
\def\refname#1{\xdef#1{\the\refno}}




\Title{\vbox{\baselineskip12pt
\hbox{
}}}
{\vbox {\centerline{ The Painlev\'e analysis for $N=2$ super KdV
equations }}}
\smallskip
\centerline{S. Bourque and P. Mathieu }

\smallskip\centerline{ \it D\'epartement de Physique,}
\smallskip\centerline{{\it Universit\'e Laval,}}
\smallskip\centerline{{\it Qu\'ebec, Canada G1K 7P4}}
\smallskip\centerline{sbourque@phy.ulaval.ca}
\smallskip\centerline{pmathieu@phy.ulaval.ca}

\vskip .2in
\bigskip
\bigskip
\centerline{\bf Abstract}
\bigskip
\noindent

The Painlev\'e analysis of a generic multiparameter $N=2$ extension of the
Korteweg-de Vries equation is presented. Unusual aspects of the
analysis, pertaining to the presence of two fermionic fields, are
emphasized. For the general class of models considered, we find that
the only ones which manifestly pass the test are precisely the four
known integrable supersymmetric KdV equations, including the SKdV$_1$ case.

\leftskip=0cm \rightskip=0cm

\Date{06/00 (revised 02/01)\ }
\let\n\noindent


\newsec{Motivation and summary of the results}

Even in its simplest form, the Painlev\'e analysis [\ref{J. Weiss, M. Tabor
and G. Carnavale, J. Math. Phys. {\bf 24} (1983) 522.}], as applied to a
multiparameter class of equations, is a powerful tool for identifying those
special values of the parameters for which the equations are potentially
integrable.

However, for fermionic -- and in particular, supersymmetric -- extensions of
known integrable systems, the application of the test is made a little
tricky by the presence of fermionic fields [\ref{P. Mathieu, Phys. Lett.
{\bf A128} (1988) 169}\refname\MP, \ref{L. Hlavat\'y, Phys. Lett. {\bf A137}
(1989) 173.}\refname\hla] and actually very few systems have been fully
studied so far (see also [\ref{A. Das, W-J, Huang and S. Roy, Phys. Lett.
{\bf A157} (1991) 113.}]). For a single fermionic extension (with a
fermionic field of degree $3/2$ in the normalization where the degree of
$\d_x$ is 1) of the Korteweg-de Vries (KdV) equation, which contains 3 free
parameters, the test has selected the very two integrable nontrivial
extensions of the KdV equation, namely the Kuper-KdV [\ref{B. Kupershmidt,
Phys. Lett. {\bf A 102} (1984) 213.}\refname\Ku]
(which is not invariant under supersymmetry)
and the supersymmetric KdV (sKdV - the small {\bf s}
refers to $N=1$) equation [\ref{P. Mathieu, J. Math. Phys. {\bf 29} (1988)
2499}\refname\Mat, \ref{P. Mathieu, Phys. Lett. {\bf B203} (1988)
287.}\refname\Mata, \ref{Yu. I. Manin and A.O. Radul, Comm. Math, Phys. {\bf
98} (1985) 65.}\refname\MR] - the latter being called here the sKdV$_3$
equation for reasons explained below. There is an additional integrable
supersymmetric system [\Mat,\ref{K. Becker and M. Becker, Mod. Phy. Lett.
{\bf A8} (1993) 1205.}] which will be referred to, in the following, as the
sKdV$_0$ equation. Although the latter is somewhat trivial in that the
fermionic fields do not appear in the bosonic evolution equations (and for
this reason it was excluded from the generic family considered in [\MP]),
this will not be an issue here [\ref{Still, another integrable system has
been found to satisfy the Painlev\'e test [\hla]. However, it is also
rather trivial (cf. its precise form given below) and thereby easily
proved to be integrable directly (see also D.-G. Zhang and B.-Z. Li,
Phys. Lett. {\bf A171} (1992) 43.). But most importantly for us, it
is not supersymmetric.}]. That the integrability of these models
had already been established by other means supports the validity
of the application of the test, or more precisely, its reliability
as an integrability indicator, in the presence of fermionic fields.

No similar studies have been performed for the extension of the KdV equation
with two fermions and an additional bosonic field. Six such systems are
known to be integrable: the three usual $N=2$ supersymmetric KdV (SKdV - the
capital {\bf S} is used for $N=2$) equations, i.e., the SKdV$_a$ equation
(where $a$ is a free parameter in a second hamiltonian formulation) for
$a=-2,1,4$ [\ref{C.A. Laberge and P. Mathieu, Phys. Lett. {\bf B215} (1988)
718-722}\refname\LaP, \ref{P. Labelle and P. Mathieu, J. Math. Phys. {\bf
32} (1991) 923-927.}\refname\Lab, \ref{ Z. Popowicz, Phys. Lett. {\bf A174}
(1993) 411.}\refname\Po], the SKdV$_{\cal O}$ (where the subscript stands
for `odd') equation [\ref{Z. Popowicz, Phys. Lett. {\bf B459} (1999)
150.}\refname\Popo], which has an {\it odd} Poisson bracket formulation, the
SKdV-B equation [\ref{A. Das and Z. Popowicz, {\it New nonlocal charges in
susy integrable models}, nlin.SI/0004034.}] and the $osp(2,2)$ KdV equation,
the direct extension of the Kuper-KdV equation (which is thus not invariant
under 
$N=2$ supersymmetry) [\LaP, \ref{ P. Mathieu and M. Thibeault, Lett. Math.
Phys. {\bf 18} (1989) 9.}].

The details of the Painlev\'e analysis of these systems has never been
presented in the literature. Actually, it has been claimed that for the
SKdV$_1$ equation, the test is failed (see in particular the concluding
remarks in [\Po]). The particular interest for this case, at the time, was
due to its conjectural integrability status for some years before the
discovery of its Lax formulation in [\Po]. But given that this system is now
known to be integrable [\ref{In addition, the integrability of the bosonic-core
version of this system is demonstrated in P. Kersten and J. Krasil'shchik, {\it
Complete integrability of the coupled KdV-mKdV system}, nlin.SI/0010041. These
authors prove the integrability by displaying an infinite series of
symmetries.}], that it does not have the Painlev\'e property sounds as an
extremely surprising statement. Clearly, the failure of the Painlev\'e test is
not by itself a clear indication of nonintegrability. For instance, the equation
might have to be somewhat transformed in order to successfully pass the test.
However, in a multiparameter deformation of an equation, we definitely expect
that if the test is satisfied for some values of the parameters
(corresponding to a known integrable system), it should be equally
satisfied for all other values of the parameters for which the equations
are known to be integrable. But to rule on the SKdV$_1$ equation, we need
to perform the test for the other cases too in order to see if, in the
presence of two fermionic fields, it is again a reliable integrability
indicator.

The natural expectation is that all six extended KdV equations known to be
integrable should have the Painlev\'e property. However, precisely because
there are two fermions, the test displays unusual features. This point in
itself is certainly not surprising given some of the odd technical aspects
of the test as applied to a single fermionic extension of the KdV equation
[\MP, \hla]. Clearing up the status of the Painlev\'e property for the
SKdV$_1$ equation was our first motivation for this work.

We present here the result of a `complete' Painlev\'e analysis for four
supersymmetric integrable systems (excluding the SKdV-B equation by
requiring an $O(2)$ invariance -- see below). More precisely, we perform a
simplified analysis, in which, in addition to verifying the plain properties
of a genuine pole behavior of the leading singularities and the
integrality of the resonance positions, we only check the
compatibility conditions at the nonnegative resonances. The qualitative
`complete' refers to the fact that we consider the full set of
four evolution equations in each case. In addition to be rather
complicated, even though the analysis is done with the simplified
Kruskal ansatz [\ref{M. Jimbo, M.D. Kruskal and T. Miwa, Phys. Lett.
{\bf A92} (1982) 52.}], it reveals an unusual feature: in two cases
out of four (and this includes the SKdV$_1$ equation), in order to
verify the last resonance conditions -- whenever this resonance is
bosonic -- say, at level $n$, we need to solve the set of recursion
equations at level $n+1$. In other words, at first sight the
compatibility conditions are {\it not} satisfied. However, they
involved some field components that get determined only at the next
recursion level. But when this is done and the solutions are
substituted back into the level $n$ resonance relations, the
compatibility conditions are found to be satisfied. We thus conclude that,
in this context, the Painlev\'e test is still in par with the other
integrability indicators.

A second motivation for this work was to initiate the search for new
integrable $N=2$ extensions of the KdV equations by using the Painlev\'e
property as a probing tool to test generic deformations of the known SKdV
equations. In the present work we treat the most general deformation (which
contains 4 free parameters) compatible with a natural $O(2)$ invariance.

Instead of starting with a brute force analysis of this four-parameter
equation, we use a simple observation in order to constrain these
parameters, which is that the reduction (by which we mean setting some
fields equal to zero) of an integrable system has to be integrable. For
instance, a clear signal of this integrability persistence is that, after
the reduction of an integrable system, there remains an infinite number of
conservation laws. In particular, the $N=1$ reduction of an integrable SKdV
equation has to be either the sKdV$_3$ or sKdV$_0$ equations. This fixes two
parameters and selects two classes of two-parameter equations. Another
simplifying feature of the above observation is that the bosonic core of the
full set of equations (obtained by setting the two fermionic fields equal to
zero) must also be integrable. The analysis of such bosonic systems (here a
system of two coupled evolution equations) is much easier and puts severe
constraints on the remaining parameters. In fact, the bosonic core of the
test is satisfied (modulo a technical restriction discussed below) for
only four cases, which are precisely {\it the four known integrable
supersymmetric systems}.

Our search for new systems is thus unsuccessful. The results suggest in
particular, that (most probably) there  are no integrable deformations
of the SKdV$_{\cal O}$ equation.

The analysis of the complete fermionic systems is then performed case by
case and the Painlev\'e property is verified in all four cases, as already
mentioned.

We should point out a technical limitation of the present analysis, which
is restricted to the study of the so-called {\it principal family} -- in
the terminology of [\ref{R. Conte, A.P. Fordy and A. Pickering, Physica D
{\bf 69} (1993) 33.}\refname\Conte]. That means that we only look for
nonnegative resonances, in addition to the resonance at level $n=-1$.
For a complete analysis, 
solutions with 
negative resonances  
must also be considered. The perturbative Painlev\'e
test [\Conte] provides a method for investigating such
solutions. 
However, there is no finite algorithm ensuring the absence of movable
logarithms (only for the principal families one can guarantee that the
system has the Painlev\'e property). As a result, the computations are
much more involved. We intend to return to this question elsewhere. Here
we only indicate the cases where the negative resonances occur but without
further analysis. Our statements concerning the non-existence of new
integrable systems must thereby be tinged by this technical restriction.

The article is organized as follows. In section 2, we present the
general class of $N=2$ supersymmetric equations to be studied and
discuss the constraints resulting from integrability under
truncation to $N=1$ supersymmetric equations. The general
structure of the recursion relations is displayed in section 3.
The delicate question of fixing the dominant resonance of the
fermionic fields is discussed in full detail in appendix B. In
the following section, we present the essential results of the
bosonic-core analysis, relegating the details to appendix A. This
analysis turns out to be rather involved, necessitating the
consideration of a large number of special cases. Finally,
section 5 presents a brief discussion of the study of the
equations incorporating the fermionic fields. Here we only
present the salient features of the SKdV$_1$ case and briefly
comment on the differences that occur in the other cases. Our
conclusions are reported in section 6.


\newsec{The general equations and the $N=1$ constraints}

The $N=1$ supersymmetrization of the KdV equation
$$ u_t= -u_{xxx}+6uu_x\eq$$
is obtained by extending the $u$ field to a fermionic superfield as
$$u(x)\quad \rw \quad \phi(x,\ta) = \ta u(x)+\xi(x).\eq$$
Here $\ta$ is a grassmannian variable ($\ta^2=0$) and $\xi$ is a fermionic
field: $\xi(x)\xi(x')=-\xi(x')\xi(x)$. The direct supersymmetrization reads
[\Mat,\ref{See also P. Mathieu, {\it Open problems for SuperKdV
equations}, math-ph/0005007, for a pedagogical introduction to the
supersymmetric KdV equations.}]
$$\phi_t= -\phi_{xxx}+c\,(\phi
D\phi)_x+(6-2c)\,\phi_x(D\phi)\,,\eqlabel\skdv$$
where $c$ is a free parameter and $D$ is the superderivative:
$D=\ta\d_x+\d_\ta$ so that $D^2=\d_x$. It turns out that this
equation is integrable only if $c=0$ or $3$ [\Mat].
We call the resulting equation the sKdV$_c$
equation. Its component version reads
$$\eqalign{
&u_t= -u_{xxx}+6\,uu_x- c\,\xi\xi_{xx}\cr
&\xi_t= -\xi_{xxx}+(6-c)\,u\xi_x + c\,u_x\xi\,.\cr}\eqlabel\skdvco$$
For $c=0$ we see that $\xi$ decouples from the first equation [\ref{Notice
that in [\hla], the following system has also been shown to have the
Painlev\'e
property:
$$
u_t= -u_{xxx}+6uu_x- 3\xi\xi_{xx}\qquad\quad
\xi_t= -\xi_{xxx}+6(u\xi)_x\,.$$
However, the change of variable $v=u-\xi\d_x^{-1}\xi$ transforms it
into (cf. D. Depireux and P. Mathieu, Phys. Lett. {\bf B 308} (1993)
272):
$$v_t= -v_{xxx}+6vv_x\qquad\quad
\xi_t= -\xi_{xxx}+6(v\xi)_x$$
which is somewhat trivial (but clearly integrable) and manifestly not
supersymmetric invariant.}].

The $N=2$ super-extension is obtained by lifting $u$ to a bosonic superfield
defined as follows (with the time dependence being implicit):
$$\Phi(x,\ta_1,\ta_2)= \ta_2\ta_1 u(x) +
\ta_1\xi^{(2)}(x)+\ta_2\xi^{(1)}(x)+ w(x)\, ,\eq$$
$\xi^{(1)}$ and $\xi^{(2)}$ are two fermionic fields and $w$ is a new
bosonic field. Using the superderivatives
$$D_i= \ta_i\d_x+\d_{\ta_i}\qquad \Rw \qquad D_i^2=\d_x \quad (i=1,2) \qquad
D_1D_2=-D_2D_1\eq$$
the most general version (subject to some restrictions to be specified
shortly) of the $N=2$ extension of the KdV equation reads
$$\Phi_t = -\Phi_{xxx}+\a_1\,\Phi D_1D_2\Phi_x +\a_2\,\Phi_x D_1
D_2\Phi+{\a_3\over 2}\,(D_1D_2\Phi^2)_x+{\a_4 \over 3}\,(\Phi^3)_x\, .\eq$$
This equation contains all possible terms that are compatible with an
homogeneity requirement under a gradation defined by deg $\Phi =$ 1, deg
$D_i=1/2$ and the $O(2)$ invariance, that is, invariance under the
transformation $\Phi \rw -\Phi$ and $D_1 \lra D_2$. (For instance, terms like
$[D_1(\Phi_xD_2\Phi)- D_2(\Phi_xD_1\Phi)]$ or $[D_1(\Phi D_2\Phi_x)-
D_2(\Phi D_1\Phi_x)]$ or even $[(D_1\Phi)(D_2\Phi_x)-(D_2\Phi)(D_1\Phi_x)]$
are not independent -- i.e., they are linear combinations of those already
given).

The $N=1$ reduction is obtained by setting
$$\Phi(x,\ta_1,\ta_2)=\ta_2 \phi(x,\ta_1)+{\cal F}(x,\ta_1)\eq$$
and keeping only the linear terms in $\ta_2$ with ${\cal F}=0$. All {\it
integrable} versions of this four-parameter equation {\it must reduce} to
the sKdV$_c$ equation for either $c=0$ or 3. This fixes two
parameters:
$$\a_1=c\;,\qquad \a_2=6-c\,. \eq$$
The other two are redefined as follows:
$$\a_3=\a -1\;,\qquad \a_4=\beta\eq$$
and we are left with two distinct two-parameter equations:
$$\eqalign{
\Phi_t &= -\Phi_{xxx}+c\,\Phi D_1D_2\Phi_x +(6-c)\,\Phi_x
D_1D_2\Phi+{\a-1\over 2}\,(D_1D_2\Phi^2)_x\cr
&\quad+\beta\,\Phi^2\Phi_x}\eq$$
($c=0,3$).
In terms of component fields, it leads to four coupled equations:
$$\eqalign{
u_t&=-u_{xxx}+6\,uu_x-c\,\xi^{(i)}\xi^{(i)}_{xx}-c\,ww_{xxx}
-(6-c)\,w_xw_{xx}\cr
&\quad-{\a-1\over 2}\,(w^2)_{xxx}+\beta\,(uw^2)_x
+2\beta(\xi^{(2)}\,\xi^{(1)}w)_x\cr
\xi^{(i)}_t &=-\xi^{(i)}_{xxx}+c\,u_x \xi^{(i)}
+(6-c)\,u\xi^{(i)}_x-c\,\e_{ij}\xi^{(j)}_{xx}w\cr
&\quad-(6-c)\,\e_{ij}\xi^{(j)}_xw_x
-(\a-1)\,\e_{ij}(\xi^{(j)} w)_{xx}+\beta\,(\xi^{(i)}w^2)_x\cr
w_t&=-w_{xxx}+c\,u_xw+(6-c)\,uw_x
+(\a-1)\,(uw+\xi^{(2)}\xi^{(1)})_x\cr
&\quad+\beta\, w^2w_x\cr}\eqlabel\SKdV$$
with $i,j=1,2$ and $\e_{12}=-\e_{21}=1$.


\newsec{The Painlev\'e analysis: recursion relations }

We next proceed with the Painlev\'e analysis by solving the recursion
equations in order to find those values of the parameters $\a$ and $\beta$
for which the test is satisfied. In the present work, we content ourself
with a minimal version of the test, which consists in verifying:

\n 1- that the leading singularity is integer (i.e., pole-like),

\n 2- the resonances occur at integer levels,

\n 3- the compatibility conditions are satisfied at the nonnegative
resonances.

\n We will further give all possible solutions with integer resonances
but without further analysis of these last cases.

The expansion of the component fields about a movable singular manifold
$\vphi(x,t)$ reads
$$u=\sum_{n=0}^\infty{u_n\vphi^{n-p}}\,,
\qquad \xi^{(i)}=\sum_{n=0}^\infty{\xi^{(i)}_n\vphi^{n-r}}\,,
\qquad w=\sum_{n=0}^\infty{w_n\vphi^{n-q}}\eqlabel\series$$
\n $(i=1,2)$. By symmetry, the value of the leading singularity must be the
same for the two fermionic fields. To simplify the analysis, we will use the
Kruskal's ansatz:
$$\vphi(x,t)=x-f(t)\,, \qquad u_n=u_n(t)\,, \qquad
\xi^{(i)}_n=\xi^{(i)}_n(t)\,,
\qquad w_n=w_n(t)\,. \eqlabel\Kansatz$$
The first step amounts to fix the leading singularity: we easily find that
$p=2,\; q=1$. Note that this is a consequence of the SKdV
degree-homogeneity already mentioned: setting deg$(\d_x)=1$,
it follows that deg$(u)=2$ and deg$(w)=1$. Now since deg$(\vphi)=$
deg$(x)=-1$ and
$u_0$ and $w_0$ are constants, hence of degree zero, we thus conclude that
deg$(u)=2$ and deg$(w)=1$ only if $p=2$ and $q=1$.

The determination of the leading singularity for the fermionic
fields is a bit tricky (see for instance [\hla]); it is shown in
appendix B that $r=2$ is a solution and
all other possible solutions do not
pertain to a principal family. However, for the rest of this
section, we leave $r$ unspecified since the recursion relations
themselves are needed to fix it -- cf. appendix B. Moreover, the
precise value found for $r$ depends explicitly on the first
bosonic terms and these are fixed from the bosonic-core analysis.

A direct substitution of (\series), (\Kansatz) with $p=2,\; q=1$ into
(\SKdV) leads to the general recursion formulae:
\xdef\rcuf{\secsym\the\meqno}
$$\eqalignno{
u_{n-3,t}+(n-4)\,&u_{n-2}\vphi_t\cr
&=-(n-2)(n-3)(n-4)u_n+3\,(n-4)\sum_{m=0}^n{u_{n-m}u_{m}}\cr
&\quad -(\a-1+c)\sum_{m=0}^n{(m-1)(m-2)(m-3)\,w_{n-m}w_{m}}\cr
&\quad -{1 \over 2}(3\a+3-c)(n-4)\sum_{m=0}^n{(n-m-1)(m-1)\,w_{n-m}w_{m}}\cr
&\quad +\beta\,(n-4)\sum_{m=0}^n{\sum_{l=0}^m{u_{n-m}w_{m-l}w_{l}}}\cr
&\quad +{1 \over
2}c\,(n-4)\sum_{m=0}^{n+2r-3}{(n+2r-3-2m)\,\xi^{(i)}_{n+2r-3-m}\xi^{(i)}_{m}
}\cr
&\quad
+2\,\beta\,(n-4)\sum_{m=0}^{n+2r-3}{\sum_{l=0}^m{\xi^{(2)}_{n+2r-3-m}\xi^{(1
)}
_{m-l}w_{l}}}\cr
\xi^{(i)}_{n-3,t}+(n-r&-2)\,\xi^{(i)}_{n-2}\vphi_t\cr
&=-(n-r)(n-r-1)(n-r-2)\,\xi^{(i)}_n \cr
&\quad+c\sum_{m=0}^n{(n-m-2)\,u_{n-m}\xi^{(i)}_{m}}\cr
&\quad+(6-c)\sum_{m=0}^n{(m-r)\,u_{n-m}\xi^{(i)}_{m}}&(\rcuf)\cr
&\quad
+\beta\,(n-r-2)\sum_{m=0}^n{\sum_{l=0}^m{\xi^{(i)}_{n-m}w_{m-l}w_{l}}}\cr
&\quad -c\sum_{m=0}^n{(n-m-r)(n-m-r-1)\,\e_{ij}\xi^{(j)}_{n-m}w_{m}}\cr
&\quad -(6-c)\sum_{m=0}^n{(n-m-r)(m-1)\,\e_{ij}\xi^{(j)}_{n-m}w_{m}}\cr
&\quad -(\a-1)(n-r-1)(n-r-2)\sum_{m=0}^n{\e_{ij}\xi^{(j)}_{n-m}w_{m}}\cr
w_{n-3,t}+(n-3)\,&w_{n-2}\vphi_t\cr
&=-(n-1)(n-2)(n-3)\,w_n\cr
&\quad+c\sum_{m=0}^n{(n-m-2)\,u_{n-m}w_{m}}\cr
&\quad +(6-c)\sum_{m=0}^n{(m-1)\,u_{n-m}w_{m}}\cr
&\quad +(\a-1)(n-3)\sum_{m=0}^n{u_{n-m}w_{m}}\cr
&\quad +{1 \over
3}\beta\,(n-3)\sum_{m=0}^n{\sum_{l=0}^m{w_{n-m}w_{m-l}w_{l}}}\cr
&\quad
+(\a-1)(n-3)\sum_{m=0}^{n+2r-3}{\xi^{(2)}_{n+2r-3-m}\xi^{(1)}_m}\cr}$$
\n Here $n$ takes any integer value from min$(3-2r,0)$ to $\infty$
(sticking to the principal family). It is understood that every field-component
with a negative index is zero. {}For the system of equations to be integrable,
the solution needs to contain
a sufficient number of arbitrary functions. For the case under study, the
system being composed of four coupled third order equations, there should be
twelve arbitrary functions, six bosonic and six fermionic. With the leading
singularities fixed, we need to determine those (recursion) levels $n$ --
the resonances -- in (\rcuf) for which there are arbitrary functions. That
clearly requires $n$ to be an integer. At each such level, the
equation must vanish identically without enforcing any
constraints on the lower-order arbitrary functions. These are the
compatibility conditions at the resonances. We then proceed in
two steps. We first find all the possible values of the free
parameters for which the bosonic-core system has the Painlev\'e
property. Then, for those special parameters, we complete the
analysis for the full system with the fermionic fields
reintroduced.


\newsec{The Painlev\'e analysis of the bosonic core}

The bosonic-core analysis is the most important and also the most involved
part of this work. It amounts to consider one-by-one a long sequence of
special cases. Although the analysis is straightforward for most of them,
there is a number of cases (that include cases in which the test is
satisfied) for which this is not so. For this reason, a somewhat detailed
presentation of all the possibilities is required. It is reported in
appendix A. For the ease of reading, we collect in this section the final
results of this appendix.

The only cases for which the Painlev\'e property of the bosonic
core is fully satisfied are listed below. Note that the `body'
(i.e., without the nilpotent part) values of $u_0$ and $w_0$
represent an important part of the data since it is necessary to
fix uniquely the leading singularity of the fermionic fields.
Here,  $k=\pm i$.

\settabs\+ \indent\indent & \indent\indent\quad & \indent\indent\quad &
\indent\indent\quad & \indent\indent\quad & \indent\indent\quad &\cr

\+ (I) & $SKdV_{-2}$ \cr
\+ & $c=3\,,$ & $\a=-2\,,$ & $\beta=-6\,,$ & $u_0=-1\,,$ & $w_0=k\,.$ \cr

\+ (II) & $SKdV_1$ \cr
\+ & $c=3\,,$ & $\a=\,1\,,$ & $\beta=\,3\,,$ & $u_0=\,1\,,$ & $w_0=k\,.$ \cr

\+ (III) & $SKdV_4$ \cr
\+ & $c=3\,,$ & $\a=\;4\,,$ & $\beta=12\,,$ & $u_0={1 \over 2}\,,$
 & $w_0={1 \over 2}k\,.$ \cr

\+ (IV) & $SKdV_{\cal O}$ \cr
\+ & $c=0\,,$ & $\a=\,1\,,$ & $\beta=\,0\,,$ & $u_0=\,1\,,$ & $w_0=k\,.$ \cr

\+ (V) & $SKdV_{-2}$ (`degenerate' case) \cr
\+ & $c=3\,,$ & $\a=-2\,,$ & $\beta=-6\,,$ & $u_0=2\,,$ & $w_0=0\,.$ \cr

Cases (I)-(V) are those with nonnegative resonances. For
completeness, we also present all other possible solutions that
are not in principal families. These are listed below. It is
always understood that  $j_1$, $k_1$ and $k_2$ are integers.

\+ (VI)
 & $c=3\,,$ & $\a={1 \over 2}j_1(j_1-3)-1\,,$ && $\beta={3 \over
2}j_1(j_1-3)-1\,,$ \cr
\+  && $u_0=2\,,$ & $w_0=0$ \cr \+ & ($j_1 \geq 4$). \cr
$\,$

\+ (VII)
 & $c=3\,,$ & $\a={6 k_1 \over k_1(3-k_2)+k_2(k_2+1)-6}-2\,,$ \cr
\+  && $\beta=108 {k_1 - 2 \over (k_1(3-k_2)+k_2(k_2+1)-6)^2}\,,$ \cr
\+  && $u_0={1 \over 6} (k_1(3-k_2)+k_2(k_2+1)-6)\,,$ \cr
\+  && $w_0=k\,u_0$ \cr
\+ & ($k_1 \ge \max\,(5, 2k_2+1)\,, \qquad k_2 \ge -1$). \cr
$\,$

\+ (VIII)
 & $c=3\,,$ & $\a={6 k_1 \over k_1(3-k_2)+k_2(k_2+1)-6}-2\,,$ \cr
\+  && $\beta=108 {k_1 - 2 \over (k_1(3-k_2)+k_2(k_2+1)-6)^2}\,,$ \cr
\+  && $u_0={1 \over 6} (k_1(3-k_2)+k_2(k_2+1)-6)\,,$ \cr
\+  && $w_0=k\,u_0$ \cr
\+ & ($k_1 \ge 5\,, \qquad k_2 \le -4 $). \cr
$\,$

\+ (IX)
 & $c=3\,,$ & $\a=2{7-k_1^2 \over 2+k_1^2}\,,$ &&
$\beta={108 \over (2+k_1^2)^2}\,,$ \cr
\+  && $u_0={1 \over
6}(2+k_1^2)\,,$ && $w_0=k\,u_0$ \cr
\+ & ($k_1 \ge 5$). \cr
$\,$

\+ (X)
 & $c=3\,,$ & $\a=-4{k_1 \over 5k_1+6}\,,$
&& $\beta=108 {k_1 \over (5 k_1+6)^2}\,,$ \cr
\+  && $u_0={5 \over
6}k_1 + 1\,,$ && $w_0=k\,u_0$ \cr
\+ & ($k_1 \le -7$). \cr
$\,$

\+ (XI)
 & $c=3\,,$ & $\a=-4{k_1 \over 5k_1+6}\,,$
&& $\beta=108{k_1 \over (5 k_1+6)^2}\,,$ \cr
\+  && $u_0={5 \over
6}k_1 + 1\,,$ && $w_0=k\,u_0$ \cr
\+ & ($k_1 \ge 3$). \cr
$\,$

\+ (XII)
 & $c=0\,,$ & $\a=7\,,$ &
$u_0=2\,,$ & $w_0=0$.\cr
$\,$

\+ (XIII)
 & $c=0\,,$ & $\a=1\,,$ & $u_0=2\,,$ & $w_0=0$.\cr
$\,$

\+ (XIV)
 & $c=0\,,$ & $\a=-{7 \over 31}\,,$ & $\beta=-{78 \over 961}\,,$ &
$u_0=-31\,,$ & $w_0=k\,u_0$.\cr
$\,$

\+ (XV)
 & $c=0\,,$ & $\a=-{5 \over 21}\,,$ & $\beta=-{6 \over 49}\,,$ &
$u_0=-21\,,$ & $w_0=k\,u_0$.\cr
$\,$

\+ (XVI)
 & $c=0\,,$ & $\a=-{1 \over 7}\,,$ & $\beta={15 \over 98}\,,$ &
$u_0=14\,,$ & $w_0=k\,u_0$.\cr
$\,$

\+ (XVII)
 & $c=0\,,$ & $\a=-{7 \over 5}\,,$ &
$\beta=0\,,$ & $u_0=-5\,,$ & $w_0=k\,u_0$.\cr
$\,$


\newsec{The Painlev\'e analysis of the full fermionic systems of the four
integrable $N=2$ supersymmetric systems}

In a second step, the Painlev\'e analysis is completed for the fermionic
extension of the successful bosonic systems. We omit the
details of the SKdV$_{-2,\,4,\,{\cal O}}$ analysis
and sketch
some aspects of the analysis of the SKdV$_1$ equation.


\subsec{Analysis of the SKdV$_1$ equation}

In appendix B, it is shown that the leading singularity of fermionic fields
must be $r=2$ and that the following condition must hold:
$$\xi^{(2)}_0=k_0\,\xi^{(1)}_0\eq$$
where $k_0^2=-1$ [\ref{Notice that even if the Painlev\'e analysis breaks
the $O(2)$ invariance, we are still free to choose the sign of $k_0$.
Clearly, in a formulation in terms of the redefined fields
$\xi^{(\pm)}=\xi^{(1)}\pm i \xi^{(2)}$, we would be free to take either
$\xi^{(+)}_0$ or $\xi^{(-)}_0$ as an arbitrary function, the other being
null.}]. With this condition, (\rcuf) for $n=-1$, which reads
$$
-30\,\a\,\xi^{(1)}_0\xi^{(2)}_0 w_0=0\qquad\quad
-4(\a-1)\,\xi^{(1)}_0\xi^{(2)}_0=0\,,\eqlabel\frstrc$$
is automatically satisfied.

{}From the resonance equations obtained in appendices A and B, the bosonic
resonances must occurs at the roots of
$$(n+1)(n-1)(n-2)(n-3)(n-4)(n-6)=0\eq$$
corresponding to the arbitrariness of $\vphi, w_1, w_2, w_3, u_4$ and $w_6$,
whereas the fermionic ones are determined by the roots of
$$n(n-2)^2(n-4)^2(n-6)=0\eq$$
corresponding to the arbitrariness of
$\xi^{(1)}_0,\;\xi^{(1)}_2,\;\xi^{(2)}_2,\;\xi^{(1)}_4,\;\xi^{(2)}_4$ and
$\xi^{(1)}_6$.

The introduction of the fermionic fields brings a little complexity right at
the beginning of the analysis in that it is necessary to use both the $n=0$
and $n=1$ conditions in order to fix $u_0$, $\xi^{(i)}_0$ and $w_0$
unambiguously. Once this is settled, the remaining part of the analysis is
straightforward, apart from the plain fact that the equations are rather
complicated.

The most general solution to the recursion formulae at level $n=0$, for
which the bosonic part reduces to the one found in the bosonic-core analysis
(with constant $k$ appearing in $w_0$ fixed to $k=-k_0$, as shown in
appendix B), is:
$$\eqalign{
u_0=&\;\;1-{2 \over 3}\lambda_0\,\xi^{(1)}_0\,,\cr
w_0=&-k_0+\lambda_0\,\xi^{(2)}_0,\cr
}\qquad\eqalign{
\xi^{(2)}_0=&\;k_0\,\xi^{(1)}_0,\cr
\xi^{(2)}_1=&\;k_0\,\xi^{(1)}_1+{4 \over 3}k_0\lambda_0 +
k_1\xi^{(2)}_0,\cr}\eqlabel\fOa$$
where $k_0^2=-1$, $k_1$ is a (even) constant, $\lambda_0$ is a fermionic
constant and $\xi^{(1)}_0$ is an arbitrary fermionic function.

In order to fix uniquely $u_0$ and $w_0$, we need to consider the equations
for $n=1$. At this level, a substitution of (\fOa) into the recursion
equations leads to $(k_0^2=-1)$:
$$\eqalign{
u_0&=1\,,\cr
u_1&=0\,,\cr
}\quad\eqalign{
\xi^{(1)}_0&\; {\rm arbitrary}\,,\cr
\xi^{(1)}_1&=0\,,\cr
}\quad\eqalign{
\xi^{(2)}_0&=k_0\xi^{(1)}_0,\cr
\xi^{(2)}_1&=0\,,\cr
}\quad\eqalign{
w_0&=-k_0\,,\cr
w_1&=\;0\,.\cr
}\eq$$

Pursuing the analysis of (\rcuf), one can verify that all the compatibility
conditions at the various resonances are satisfied. Since the resulting
equations are very long, this part of the analysis will be omitted. Notice
that since the equation for
$w_6$ depends upon the value of
$\xi^{(i)}_7$
$(i=1,2)$, we have to go up to level $n=7$ to fix completely the different
non-arbitrary functions needed to verify this particular compatibility
condition. The analysis for levels $n=5$ to $n=7$ is actually very
complicated; the computations have been made with Maple (with the package
Grassmann).


\subsec{Comments on the other three cases}

The analysis for the other three cases singled out by the bosonic-core
analysis has also been performed successfully.

{}For the SKdV$_{-2}$ equation, only one of the two possible cases
identified by the bosonic-core analysis is found to have the
Painlev\'e property: this is case (I). The analysis for this case
is not too difficult since the last resonance is fermionic,
occurring at level $n=5$, so that we only have to push the
analysis up to this level. The arbitrary functions are:
$\vphi\,,\;\xi^{(1)}_0,\; u_1,\; \xi^{(1)}_2,\; u_3,\;
\xi^{(1)}_3,\; w_3,\; u_4,\;\xi^{(1)}_4,\; \xi^{(2)}_4,\; w_4$
and $\xi^{(1)}_5$.

{}For the so-called `degenerate' SKdV$_{-2}$ case (V), the Painlev\'e test
immediately fails at the first level since both $\xi^{(1)}_0$ and $\xi^{(2)}_0$
need to be arbitrary and, at the same time, satisfy
$\xi^{(1)}_0\xi^{(2)}_0=0$ (cf. (\frstrc)). This condition is completely
independent of the value of $r$.

{}For the SKdV$_4$ equation, the compatibility conditions are all verified
and
$\vphi$, $\xi^{(1)}_0$, $ w_1$, $\xi^{(1)}_2$, $ u_3$, $\xi^{(1)}_3$, $ w_3$,
$u_4$, $\xi^{(1)}_4$, $\xi^{(2)}_4$, $u_5$ and $\xi^{(1)}_5$ are found to be
arbitrary. Notice that there are two resonances at level $n=5$, one of which
being bosonic; the analysis must then be extended up to level $n=6$.

{}For the SKdV$_{\cal O}$ equation, the bosonic evolution equations decouple
from the fermionic ones; this eliminates the necessity of extending the
analysis to a higher level in order to check the compatibility condition at
the highest resonance. With $r=2$ we find that $\vphi\,,\; \xi^{(1)}_0,\;
w_1,\;
\xi^{(1)}_2,\;\xi^{(2)}_2,\; w_2,\;\xi^{(1)}_3,\; w_3,\; u_4,\;
\xi^{(1)}_4,\; u_6$ and $\xi^{(1)}_7$ are arbitrary hence it is not
necessary to perform the test for other values of $r$ (since we already know
that the system SKdV$_{\cal O}$ has the Painlev\'e property).


\newsec{Conclusion}

In this work, we have presented the Painlev\'e analysis (at
least, a reduced form of it) for the complete set of bosonic and
fermionic evolution equations pertaining to a general
multiparameter family of $N=2$ supersymmetric equations. Such an
analysis is interesting for a number of reasons. First, very few
fermionic extensions of integrable systems have been analyzed
from that point of view. The detailed analysis of specific
examples is of a clear interest in view of confirming (or
limiting or even, in principle, invalidating) the direct
extension of the test to fermionic systems. The successful
analysis presented here for four $N=2$ supersymmetric extensions
of the KdV equation, known to be integrable from other methods,
indeed confirms the validity of the naive extension of the test.
This, in turn, gives credit to the test when viewed as an
exploratory tool in the search for new integrable systems among a
multiparameter class of equations. In that respect, the present
results have not signaled the existence of a single new
integrable equation (although new integrable systems could still in
principle be
revealed by an analysis that goes beyond the principal families).

Manifestly, that only a rather limited number of examples have been studied
so far is partly due to the intrinsic complications of such computations
(involving here four coupled nonlinear equations); however, it is also due
to the special complications brought by the fermionic fields themselves. In
particular, even the determination of the leading singularity is somewhat
problematic (and a simple degree-homogeneity requirement cannot be put
forward). This particular question has been treated in great detail here.
Another aspect of the present analysis is to have put in
light unusual features of the verification of the compatibility
conditions for systems involving fermions, unusual in that these
conditions can be satisfied in some case only if higher-order
recursion relations are solved.

{}For completeness, we have also checked the Painlev\'e property for the
$osp(2,2)$ KdV equation, which is $O(2)$ symmetric but not supersymmetric,
and found that it successfully passes the test. The equations take the
form
$$\eqalign{
u_t&=\d_x[-u_{xx}+3u^2-12\xi^{(i)}\xi^{(i)}_x+24\xi^{(2)}\xi^{(1)}w+2(w_x)^2
+2ww_{xx}-6uw^2+3w^4]\cr
\xi^{(i)}_t&=-4\xi^{(i)}_{xxx}+3u_x\xi^{(i)}+6u\xi^{(i)}_x+6w^2\xi^{(i)}_x+6
ww_x\xi^{(i)}\cr &
 -12\e_{ij}\xi^{(j)}_{xx}w-12\e_{ij}\xi^{(j)}_xw_x-4\e_{ij}\xi^{(j)}w_{xx}+
6u\e_{ij}\xi^{(j)}w-2\e_{ij}\xi^{(j)}w^3\cr}\eq
$$
with $w_t=0$. Since the $w$-evolution equation is trivial this field has no
singularity; however, relying on the degree-homogeneity property, we have
set $w=\sum^{\infty}_{n=0}{w_n\vphi^{n-1}}$ and $w_0=0$. Moreover, the
fermionic fields are also singularity-free at the leading order; therefore,
$u$ is the only field having a leading singularity, which in itself is a
rather uncommon feature.

We stress finally that the analysis has been presented here in terms of the
component fields. Hence, we
have not taken advantage of the economical superfield formalism. In fact,
the Painlev\'e test has never been formulated in superspace. That would be a
definite progress since it is only in such a case that we could face a more
refined analysis that does not rely upon the simplified Kruskal ansatz. The
benefit of such a generalization is the ability to make contact, in the
early steps of the analysis, with Backl\"und transformations and Lax pairs
(see e.g., [\ref{J. Weiss, in {\it Painlev\'e transcendents}, ed. D. Levi
and P. Winternitz, Plenum Press 1992 and references therein.}]). We hope to
report elsewhere on this topic.


\appendix{A}{Analysis of the bosonic core}

In this appendix, we analyze the bosonic core of the generic supersymmetric
KdV equations. This boils down to the study of the recursion formulae
(\rcuf) in which we set all fermionic fields equal to zero: $\xi^{(i)}_n=0$
for $i=1,2$. Further references to (\rcuf) in this appendix are to be
understood with this restriction, which transforms this system into a set of
two coupled bosonic equations.

Before considering the general recursion equations and determining its
resonances, we first analyze the recursion formulae at levels $n=0,\;1$ for
the cases $c=0,\;3$, in order to impose as much constraints as possible
in the very early steps of the analysis. For every solution found at those
levels, we need to write the resonance equation in order to identify those
cases that are potentially Painlev\'e admissible. Note however that a single
solution to the $n=0,\;1$ relations can lead to more than one resonance
equation; the different possibilities must then be analyzed one by one. The
cases $c=0,\;3$ are studied separately.


\subsec{The $c=3$ case}

The recursion formulae (\rcuf), with $c=3$, can be written under the form
$$\eqalign{
A^1_{\;1} u_n + A^1_{\;2} w_n=F^1_n\cr
A^2_{\;1} u_n + A^2_{\;2} w_n=F^2_n\cr
}\eqlabel\recua$$
where
$$\eqalign{
A^1_{\;1}(n)&=[-(n-2)(n-3)+6\,u_0+\beta\, w_0^2](n-4)\cr
A^1_{\;2}(n)&=[-(\a+2)n^2+(5\a+4)n-6(\a+1)+2\beta\, u_0](n-4)\,w_0\cr
A^2_{\;1}(n)&=(\a+2)(n-3)\,w_0\cr
A^2_{\;2}(n)&=[-(n-1)(n-2)+(\a+2)\,u_0+\beta\, w_0^2](n-3)\cr
}\eq$$
and $F^1_n$ and $F^2_n$ are functions of $u_0,u_1,...u_{n-1}$ and
$w_0,w_1,...w_{n-1}$. There is a resonance when this system is not defined,
that is, when
$$A(n)=\det|A^i_{\;j}(n)| =0 \qquad (i,j=1,2)\,. \eq$$

The substitution of $\a$, $\beta$, $u_0$ and $w_0$ (whenever they are known)
for each case identified will then yield the values of the resonance levels
$n$. It should be stressed that we are particularly interested in cases in
which
there is a resonance at level $n=-1$ (corresponding to the arbitrariness of
the
singular manifold $\vphi$) with the other ones being integers $\geq 1$,
unless either $u_0$ or $w_0$ is arbitrary, in which case we also
need a resonance at level zero. The cases for which negative resonances
appear will be given, but the search for movable logarithms will be omitted.

Here are the solutions of the recursion equations for levels $n=0\,,\,1$,
the results of the resonance analysis and their compatibility conditions.
When the test is not satisfied, we simply indicate the reason (and avoid
repeating: therefore the test is not satisfied).

\+ (i) & $u_0=0$, &&& $u_1=0$,\cr
\+ & $w_0=0$, &&& $w_1$ arbitrary.\cr
\n This case can readily be eliminated given the absence of singularities.
Similarly, cases (ii) and (iii) below could have been eliminated from the
start since there are no singularities for the field $w$; however, being
interested in a supersymmetric extension for the field $u$ -- which is thus
the `leading' field -- this restriction will not be imposed.

\+ (ii) & $u_0=2$, &&& $u_1=0$,\cr \+ & $w_0=0$, &&& $w_1=0$,\cr
\+ & $A(n)=(n+1)(n-3)(n-4)(n-6)(n^2-3n-2(\a+1))$.\cr
\n Given that the two roots of the second order polynomial are $j_1$ and
$j_2$, we thus have
$$
j_2 = 3 - j_1 \;,\qquad\quad \a = {1 \over 2}j_1(j_1-3)-1 \eq$$
and we can choose $j_1 \ge j_2$. Now since the coefficients at level 0 and 1
are fixed, a resonance at one of those levels would signal the presence of a
movable logarithm. In consequence, there is no solution in a principal
family (with both $j_1$ and $j_2 \geq 0$) free from movable logarithms. The
only other cases left are those for which
$j_1 \geq 4$.
For those cases, the compatibility conditions at level 6 are satisfied only
for $\beta=3\a$.

\+ (iii) & $\a=-2$,\cr
\+ & $u_0=2$, &&& $u_1=0$,\cr
\+ & $w_0=0$, &&& $w_1$ arbitrary,\cr
\+ & $A(n)=(n+1)(n-1)(n-2)(n-3)(n-4)(n-6)$.\cr
\n The resonances correspond to the arbitrariness of
$\vphi,\,w_1\,w_2,\,w_3,\,u_4,\,u_6$. All compatibility conditions are
verified without constraints on the parameters, except the one at level 6
which forces $\beta=3\a=-6$. This will turn out to correspond to a
non-integrable solution of the recursion relations associated to the
SKdV$_{\a=-2}$ equation.

\+ (iv) & $\a=-1$,\cr
\+ & $u_0$ not fixed yet or arbitrary, &&& $u_1=0$,\cr
\+ & $w_0=k\,\sqrt{{3 \over \beta}(u_0-2)}$, &&& $w_1=0$,\cr
\+ & $A(n)=(n+1)n(n-3)(n-4)(n^2-9n-(u_0-20)-{3 \over \beta}(u_0-2))$\cr
\n where (here and below) $k^2=-1$. The resonance at level 0 would signal a
movable logarithm if $u_0$ would
have to
be fixed. Moreover, $A(n)$ would need to
be independent of $u_0$ for this coefficient to be arbitrary. This leads to
$\beta=-3$ and
$$A(n)=(n+1)n(n-3)^2(n-4)(n-6).\eq$$
However, the compatibility conditions at level 3 are not satisfied.

\+ (v) & $\a=-1$, &&& $\beta=3[{u_0-2 \over P}]$,\cr
\+ & $u_0$ fixed by value of $\beta$, &&& $u_1=0$,\cr
\+ & $w_0=k\,\sqrt{P}$, &&& $w_1=0$,\cr
\+ & $A(n)=(n+1)n(n-3)(n-4)[(n-1)(n-8)-3\,u_0\,(u_0-2)]$\cr
\n with $P=3(u_0)^2-7u_0+12$. The resonance at level 0 signals the presence
of a movable logarithm.

\+ (vi) & $\a=-1$, &&& $\beta=-{8 \over 3}$,\cr
\+ & $u_0=-{4 \over 3}$, &&& $u_1=0$,\cr
\+ & $w_0={1 \over 2}k\,\sqrt{15}$, &&& $w_1=0$,\cr
\+ & $A(n)=(n+1)n(n-3)(n-4)[n^2-9n+{211 \over 12}]$.\cr
\n $A(n)$ has non-integer roots.

\+ (vii) & $\beta=3[{(\a+2)u_0-2 \over u_0^2}]$,\cr
\+ & $u_0$ fixed by the value of $\beta$, &&& $u_1=0$,\cr
\+ & $w_0=k\,u_0$, &&& $w_1=0$,\cr
\+ & $A(n)=(n+1)(n-3)(n-4)(n-m)[n^2+(m-9)n+6(4-u_0-m)]$\cr
\n where $m=4-(\a+2)\,u_0$. Writing the two roots of the second order
polynomial as $j_1$ and $j_2$, the general solution can be written (with
$k_1$ and $k_2$ integers)
$$
m = 4 - k_1\,, \qquad
j_1 = 3 + k_2\,, \qquad
j_2 = 2 + k_1 - k_2\,, \qquad
k_1 \geq 2k_2+1
\eq$$
with
$$\eqalign{
u_0 =& {1 \over 6} (k_1(3-k_2)+k_2(k_2+1)-6)\,, \cr
\a =& {6 k_1 \over k_1(3-k_2)+k_2(k_2+1)-6}-2\,, \cr
\beta =& 108 {k_1 - 2 \over (k_1(3-k_2)+k_2(k_2+1)-6)^2}\,.
}\eq$$
The principal families are characterized by
$$ -1 \le k_2 \le 2k_2+1 \le k_1 \le 2 \eq$$
for which the possible solutions are

\vskip0.4cm
\table{}{
&\vrule#&
\strut\quad\hfill#\hfill\quad\cr
&\omit&&$m$&&$j_1$&&$j_2$&&$\a$&&$\beta$&&$u_0$&\cr
\noalign{\hrule}
&vii.a&&$2$&&$2$&&$5$&&$4$&&$0$&&${1 \over 3}$&\cr
&vii.b&&$2$&&$3$&&$4$&&$\sim {1 \over u_0}$&&$\sim {1 \over u_0^2}$&&$0$&\cr
&vii.c&&$3$&&$2$&&$4$&&$-5$&&$-27$&&$-{1 \over 3}$&\cr
&vii.d&&$3$&&$3$&&$3$&&$-4$&&$-12$&&$-{1 \over 2}$&\cr
&vii.e&&$4$&&$2$&&$3$&&$-2$&&$-6$&&$-1$&\cr
&vii.f&&$5$&&$2$&&$2$&&$-{7 \over 5}$&&$-{81 \over 25}$&&$-{5 \over 3}$&\cr
}

\n Case vii.b can be eliminated since there are no singularities (and
moreover $\a,\;\beta \to \infty$). For cases vii.a, c, d and f, the
compatibility conditions at level $n=2,\;3,\;3,\;2$ respectively are not
satisfied. For vii.e all the conditions are satisfied so that this system
passes the test ($\vphi\,,\; u_1,\; u_3,\; w_3,\; u_4$ and $w_4$ are all
arbitrary functions). It corresponds to the bosonic core of the
SKdV$_{\a=-2}$ equation.

\n The only other solutions of interest (with negative resonances) are given
by

\vskip0.4cm
\table{}{
&\vrule#&
\strut\quad\hfill#\hfill\quad\cr
&vii.g&& $k_1 \ge {\rm max}\,(5, 2k_2+1)$&& $k_2 \ge -1 $ & \cr
&vii.h&& $k_1 \ge 5$&& $k_2 \le -4 $ & \cr
}

\+ (viii) & $\beta={1 \over 3}(\a+2)^2$,\cr \+ & $u_0={3 \over
\a+2}$, &&& $u_1=0$,\cr \+ & $w_0=k\,u_0$, &&& $w_1$ arbitrary,\cr
\+ & $A(n)=(n+1)(n-1)(n-3)(n-4)(n^2-8n-6(u_0-3))$.\cr
\n We write the two roots of the quadratic term as
$$j_1 = 4 - k_1 \,,\qquad j_2=4+k_1\,,\qquad k_1 \ge 1 \eq$$
with $k_1$ an integer and
$$u_0={1 \over 6}(2+k_1^2)\,,\qquad \a=2{7-k_1^2 \over 2+k_1^2}\,.\eq$$
The principal families
(free from movable logarithms) are characterized by
$k_1=1,\,2$ so that we have

\vskip0.4cm
\table{}{ &\vrule#& \strut\quad\hfill#\hfill\quad\cr
&\omit&&$j_1$&&$j_2$&&$\a$&&$\beta$&&$u_0$&\cr \noalign{\hrule}
&viii.a&&$2$&&$6$&&$1$&&$3$&&$1$&\cr
&viii.b&&$3$&&$5$&&$4$&&$12$&&${1 \over 2}$&\cr
}

\n {}For viii.a and b, all the compatibility conditions are satisfied. Those
systems describe the bosonic core of the SKdV$_{\a=1}$ (with
$\vphi, w_1, w_2, w_3, u_4$ and $w_6$ arbitrary) and
SKdV$_{\a=4}$ (with $\vphi\,,\;w_1,\; u_3,\; w_3,\; u_4$ and
$u_5$ arbitrary) equations respectively.

\n The other possible cases are those for which $k_1 \ge 5$.

\+ (ix) & $\beta=-{9 \over 8}\a(5\a+4)$,\cr \+ & $u_0={4 \over
5\a+4}$, &&& $u_1={1 \over 2}({11\a+4 \over \a+2})k\,w_1$,\cr
\+ & $w_0=k\,u_0$, &&& $w_1$ arbitrary. \cr
\n Note however that in the singular case where $\a=-2$, $w_1=0$
and $u_1$ is arbitrary.
\+ & $A(n)=(n+1)(n-1)(n-3)(n-4)(n+{6 \over 5}u_0-{16 \over 5})(n-{6
\over 5}u_0-{24 \over 5})$.\cr

\n We can write the two roots of the last two factors as
$$j_1=2-k_1\,,\quad\quad j_2=6+k_1\eq$$
with
$$u_0={5 \over 6}k_1 + 1\,,\qquad \a=-4{k_1 \over 5k_1+6}\,.\eq$$
\n The principal families (with $-4 \le k_1 \le 0$) are thus

\vskip0.4cm
\table{}{
&\vrule#&
\strut\quad\hfill#\hfill\quad\cr
&\omit&&$j_1$&&$j_2$&&$\a$&&$\beta$&&$u_0$&\cr
\noalign{\hrule}
&ix.a&&$2$&&$6$&&$0$&&$0$&&$1$&\cr
&ix.b&&$3$&&$5$&&$4$&&$-108$&&${1 \over 6}$&\cr
&ix.c&&$4$&&$4$&&$-2$&&$-{27 \over 2}$&&$-{2 \over 3}$&\cr
&ix.d&&$5$&&$3$&&$-{4 \over 3}$&&$-4$&&$-{3 \over 2}$&\cr
&ix.e&&$6$&&$2$&&$-{8 \over 7}$&&$-{108 \over 49}$&&$-{7 \over 3}$&\cr
}

\n The resonance conditions are not met at level $n=2,\;3,\;4,\;3,\;2$
respectively. The other solutions are

\vskip0.4cm
\table{}{
&\vrule#&
\strut\quad\hfill#\hfill\quad\cr
&ix.f&&$k_1 \le -7$&\cr
&ix.g&&$k_1 \ge 3$&\cr
}

When $c=3$, there are thus only 4 cases in principal families for which the
Painlev\'e test is satisfied for the bosonic core of our multiparameter
version of the SKdV equation. Those cases correspond to (I),
(II), (III) and (V) in the list of section 4.
There is also some other cases with negative resonances. Those cases will be
classified as (VI), (VII), (VIII), (IX), (X) and (XI).


\subsec{The $c=0$ case}

The recursion formulae (\rcuf) with $c=0$ take the form (\recua) with
$$\eqalign{
A^1_{\;1}(n)&=[-(n-2)(n-3)+6u_0+\beta\,w_0^2](n-4)\cr
A^1_{\;2}(n)&=[-(\a-1)n(n+1)+6\a(n-1)+2\beta\,u_0](n-4)w_0\cr
A^2_{\;1}(n)&=[(\a-1)(n-3)-6]\,w_0\cr
A^2_{\;2}(n)&=-(n-1)(n-2)(n-3)+(\a-1)(n-3)\,u_0\cr
&\qquad+6(n-1)\,u_0+\beta\, (n-3)\,w_0^2\,.\cr
}\eq$$

The possible solutions of the resonance conditions are now listed in turn.

\+ (i) & $u_0=0$, &&& $u_1=0$,\cr
\+ & $w_0=0$, &&& $w_1$ arbitrary.\cr
\n Again, this case is eliminated due to the absence of singularity but, as
before, we will keep the cases (ii) and (iii) below even if $w$ is not
singular.


\+ (ii) & $u_0=2$, &&& $u_1=0$,\cr \+ & $w_0=0$, &&& $w_1=0$,\cr
\+ & $A(n)=(n+1)(n-4)(n-6)(n^3-6n^2+n-2\a(n-3))$.\cr
\n Writing the last factor under the form $(n-j_1)(n-j_2)(n-j_3)$, the
constants
$j_1,\,j_2,\,j_3$ must satisfy
$$j_1+j_2+j_3=6\;,\qquad j_1 j_2+(j_1+j_2)j_3=1-2\a \;,\qquad j_1 j_2
j_3=-6\a\eq$$
and we can choose $j_2 \le j_3$.The second condition requires that $\a$ be
an integer or half-integer so that the third condition allow us to choose
$j_1 = 3m$ where $m$ is an integer. This leads to the equation
$$9m(m-1)-9(m-1)+(m-1)j_2j_3=8\,.\eq$$
$m-1$ must thus be a divisor of $8$. With this last condition, a
case-by-case analysis leads to the only two possible solutions (which are
not in principal families):

\vskip0.4cm
\table{}{
&\vrule#&
\strut\quad\hfill#\hfill\quad\cr
&\omit&&$j_1$&&$j_2$&&$j_3$&&$\a$&\cr
\noalign{\hrule}
&ii.a&&$-1$&&$1$&&$6$&&$1$&\cr
&ii.b&&$-3$&&$2$&&$7$&&$7$&\cr
}

\n Case (ii.a) can be eliminated since the resonance at level $n=1$ signals
a movable logarithm.

\+ (iii) & $\a=1$, &&& \cr
\+ & $u_0=2$, &&& $u_1=0$,\cr
\+ & $w_0=0$, &&& $w_1$ arbitrary,\cr
\+ & $A(n)=(n+1)^2(n-1)(n-4)(n-6)^2$.\cr
\n This case is not a principal family solution but all positive resonances
are verified.

\+ (iv) & $\a=0$,\cr \+ & $u_0$ not fixed yet, &&& $u_1=0$,\cr
\+ & $w_0=k\sqrt{{3 \over \beta}(u_0-2)}$, &&& $w_1=0$,\cr
\+ & $A(n)=(n+1)n(n-4)[n^3-12n^2-{1 \over
\beta}((3+5\beta)u_0-47\beta-6)n$\cr
\+ &&&& $-9{(1-3\beta) \over \beta}u_0+6{(3-10\beta) \over \beta}]$ \cr
\n (Recall that $k= \pm i$). To fix the three roots of the cubic polynomial,
$u_0$ must be fixed so that the resonance at level $n=0$ signals the
presence of a movable logarithm.

\+ (v) & $\a=0$,\cr
\+ & $u_0=12({1-\beta \over 6-11\beta})$, &&& $u_1=-(3\beta+2)\sqrt{{3 \over
10}({1
\over 6-11\beta})}\,k\,w_1$,\cr
\+ & $w_0=\sqrt{30 \over 6-11\beta}\,k$, &&& $w_1$ arbitrary,\cr
\+ & $A(n)=(n+1)n(n-1)(n-4)(n^2-11n+{126-336\beta \over 6-11\beta})$.\cr
\n There is a resonance at level $n=0$ but $u_0$ and $w_0$ are
both fixed.

\+ (vi) & $\a={(\beta\, u_0-3)\,u_0+6 \over 3u_0}$,\cr
\+ & $u_0$ fixed by the value of $\a$, &&& $u_1=0$,\cr \+ & $w_0=k\,u_0$,
&&& $w_1=0$,\cr
\+ & $A(n)=(n+1)(n-4)[n^2+({1 \over 3}\beta\, u_0^2-2u_0-5)n-\beta\,
u_0^2+6]$\cr
\+ &&& $[n^2-({1 \over 3}\beta\, u_0^2-2u_0+7)n-6u_0+2\beta\, u_0^2+12]$.\cr
\n Writing
$$\beta u_0^2 = k_1\,, \qquad u_0 = {1 \over 2}k_2 + {1 \over 6}k_1$$
the roots of the two quadratic polynomials $j_1$, $j_2$, $j_3$ and $j_4$ are
the integers satisfying
$$\eqalignT{
j_1 + j_2 &= 5 + k_2\,,&\qquad j_1 j_2 &= 6 - k_1\,, \cr
j_3 + j_4 &= 7 - k_2\,,&\qquad j_3 j_4 &= 12 + k_1 - 3 k_2\,, \cr
}\eq$$
and we choose $j_1 \le j_2$ and $j_3 \le j_4$. Introducing the auxiliary
integers $k_3$ and $k_4$ such that
$$\eqalignT{
j_1 = 3 + k_2 - k_3\,, \qquad j_2 = 2 + k_3\,, \cr
j_3 = 3 + k_4 - k_2\,, \qquad j_4 = 4 - k_4\,, \cr
}\eq$$
with $2k_4 + 1 \le k_2 \le 2 k_3 - 1$. The constraints can be written as
$$\eqalign{
k_2 (k_3 + k_4 + 1) &= 0 \cr
k_3 + k_4 &= k_3^2 + k_4^2 \cr
k_4 (k_4 + k_4 k_3 - k_3^2 - 1) &= k_1 (1 + k_3 + k_4) \cr
}\eq$$
so that the only possible solution is $(k_1, k_2, k_3, k_4) = (0, 0, 1, 0)$
which correspond to
$$(j_1, j_2, j_3, j_4, u_0, \a, \beta) = (2,3,3,4,0,\sim{1 \over u_0},\sim{1
\over u_0})$$
However, the compatibility conditions at level $n=3$ are not satisfied.

\+ (vii) & $\a={3-2u_0 \over u_0}$, &&& $\beta=-3({u_0-1 \over u_0^2})$,\cr
\+ & $u_0$ fixed by $\a$ and $\beta$, &&& $u_1=0$,\cr
\+ & $w_0=k\,u_0$, &&& $w_1=0$,\cr
\+ & $A(n)=(n+1)(n-1)(n-4)(n-m)[n^2+(m-11)n-2(2m-15)]$\cr
\n with $m=3(u_0+1)$. The two roots of the quadratic polynomial are
$$j_1,j_2={8-3u_0 \over 2}\mp {1 \over 2}\sqrt{9u_0^2-8}\eq$$
Since $j_1$, $j_2$ and $m$ are integers, we must have $u_0 = {1 \over 3}
k_1$ with $k_1$ an integer. This leads to the condition
$$9u_0^2-8=k_1^2-8=k_2^2\eq$$
where $k_2$ is also an integer but this equation has a solution only when
$k_1=\pm 3$. With $k_1=-3$, there should be a resonance at level $n=0$ so
that there is a movable logarithm. The only solution is thus $u_0=1$, which
yields $m=6, \,j_1=2$, $j_2=3$ and $\a=1\,, \,\beta=0$. All the resonance
conditions are verified: $\vphi\,,\; w_1,\; w_2,\; w_3,\; u_4$ and $u_6$ are
genuine arbitrary functions. Actually, this system is the bosonic core of
the SKdV${_{\cal O}}$ equation.

\+ (viii) & $\a={1 \over 5}({4-u_0 \over u_0})$, &&& $\beta={6
\over 5}({2u_0-3 \over u_0^2})$,\cr
\+ & $u_0$ fixed by $\a$ and
$\beta$, &&& $u_1=-2{9u_0-11 \over 9u_0+4}k\,w_1$,\cr
\+ & $w_0=k\,u_0$, &&& $w_1$ arbitrary,\cr
\+ & $A(n)=(n+1)(n-1)(n-4)(n-m)[n^2-(11-m)n+2m]$\cr
\n with $m={6 \over 5}(4-u_0)$. Writing the roots of the second order
polynomial as $j_1$ and $j_2$, we thus have
$$j_1+j_2 = m-11\,,\qquad j_1 j_2 = 2m\eq$$
and we choose $j_1 \le j_2$. Elimination of $m$ leads to the formula
$$j_2={26 \over j_1+2}-2\eq$$
so that $2+j_1$ must be a divisor of $26$. We thus find that the only
possible solutions are

\vskip0.4cm
\table{}{
&\vrule#&
\strut\quad\hfill#\hfill\quad\cr
&\omit&&$m$&&$j_1$&&$j_2$&&$u_0$&&$\a$&&$\beta$&\cr
\noalign{\hrule}
&viii.a&&$ 42$&&$-28$&&$- 3$&&$-31$&&$-{7 \over 31}$&&$-{78 \over 961}$&\cr
&viii.b&&$ 30$&&$-15$&&$- 4$&&$-21$&&$-{5 \over 21}$&&$-{6 \over 49}$&\cr
&viii.c&&$  0$&&$  0$&&$ 11$&&$  4$&&$0$&&${3 \over 8}$&\cr
&viii.d&&$-12$&&$- 1$&&$ 24$&&$ 14$&&$-{1 \over 7}$&&${15 \over 98}$&\cr
}

\n Case (viii.c) can be eliminated since there are movable logarithms at
level $n=0$.

\+ (ix) & $\a={1 \over 3}$, &&& $\beta=0$,\cr
\+ & $u_0={3 \over 2}$, &&& $u_1=-{2 \over 7}k\,w_1$,\cr
\+ & $w_0=k\,u_0$, &&& $w_1$ arbitrary,\cr
\+ & $A(n)=(n+1)(n-1)(n-3)(n-4)(n^2-8n+6)$.\cr
\n $A(n)$ has non-integer roots.

\+ (x) & $\beta=0$,\cr
\+ & $u_0={2 \over \a+1}$, &&& $u_1=0$,\cr
\+ & $w_0=k\,u_0$, &&& $w_1=0$,\cr
\+ & $A(n)=(n+1)(n-3)(n-4)(n-m)(n^2+(m-9)n+6)$\cr
\n with $m=2(2-u_0)$. The roots of the quadratic piece are
$$j_1,j_2={9-m \over 2} \mp {1 \over 2}\sqrt{m^2-18m+57}\eq$$
with $j_1 \le j_2$. We can write the quantity inside the square root as
$$m^2-18m+57=(m-9)^2-24=k_1^2\eq$$
where $k_1$ must be an integer. In consequence, we must have $m=14$ or
$m=4$. The choice $m=4,j_1=2,j_2=3$ leads to $\a \sim \infty$ and
$u_0=w_0=u_1=0$ so that there are no singularities at all. The only
possible case is thus
$$(m\,, j_1\,, j_2\,, u_0\,, \a\,, \beta)=(14\,,-3\,,-2\,, -5\,, -{7 \over
5}\,, 0)\,.\eq$$

\+ (xi) & $\a=0$, &&& $\beta=0$,\cr
\+ & $u_0=2$, &&& $u_1=0$,\cr
\+ & $w_0$ arbitrary, &&& $w_1=0$,\cr
\+ & $A(n)=(n+1)n(n-4)(n^3-12n^2+w_0^2n+37n-6+3w_0^2)$.\cr
\n There is a resonance at level $n=0$ but $u_0$ is already fixed
and $w_0$ cannot be arbitrary since it enters in the expression
of the other resonances.

\+ (xii) & $\a=0$, &&& $\beta=0$,\cr
\+ & $u_0=2$, &&& $u_1$ arbitrary,\cr
\+ & $w_0=\sqrt{5}\,k$, &&& $w_1=\sqrt{5}\,k\,u_1$,\cr
\+ & $A(n)=(n+1)n(n-1)(n-4)(n^2-11n+21)$.\cr
\n There is a resonance at level $n=0$ while $u_0$ and $w_0$ are
both fixed and moreover $A(n)$ has non-integer roots.

{}For $c=0$, we have thus found only one case in a principal family for
which the
bosonic core passes the test: this is case (IV) of section 4. Some other
possibilities
can be identified as cases (XII) through (XVII).


\appendix{B}{Leading singularity and resonance equation for fermionic
fields}

The general recursion equations for the fermionic evolution equations can be
written as (cf. (\rcuf)):
$$\eqalign{
B^1_{\;1}(n)\xi^{(1)}_n+B^1_{\;2}(n)\xi^{(2)}_n&=G^1_n\cr
B^2_{\;1}(n)\xi^{(1)}_n+B^2_{\;2}(n)\xi^{(2)}_n&=G^2_n\cr
}\eqlabel\ferru$$
where
$$\eqalignD{
B^1_{\;1}(n)&=\; B^2_{\;2}(n)&=-(n-r)(n-r-1)(n-r-2)-2c\,u_0\cr
&&\qquad+(6-c)(n-r)\,u_0+\beta\,(n-r-2)\,w_0^2\cr
B^2_{\;1}(n)&=-B^1_{\;2}(n)&=[c\,(n-r)(n-r-1)-(6-c)(n-r)\cr
&&\qquad+(\a-1)(n-r-1)(n-r-2)]\,w_0\cr
}\eqlabel\ferrua$$
$r$ is the leading singularity exponent (so that $r \le 0$ corresponds to no
singularity) and $G^{\,i}_n$ $(i=1,2)$ are functions of
$\vphi,u_0,...u_n,\xi^{(1,2)}_0,...\xi^{(1,2)}_{n-1},w_0,...,w_n$.

With $n=0$, we have
$$\eqalign{
G^{\,i}_0&=0\cr
B^1_{\;1}(0)&=r(r+1)(r+2)-2c\,u_0-(6-c)r\,u_0-\beta\,(r+2)\,w_0^2\cr
B^2_{\;1}(0)&=[c\,r(r+1)+(6-c)r+(\a-1)(r+1)(r+2)]\,w_0.\cr
}\eqlabel\fsing$$
Multiplying the first equation in (\ferru) with $n=0$ by $\xi^{(1)}_0$ and
the second by $\xi^{(2)}_0$ yields (using $G^{\,i}_0=0$)
$$B^1_{\;1}(0)\xi^{(1)}_0\xi^{(2)}_0=
B^2_{\;1}(0)\xi^{(1)}_0\xi^{(2)}_0=0\eqlabel\sery$$
There are thus two possible types of solutions: either the two $B^i_{\;j}$
coefficients vanish or $\xi^{(1)}_0\xi^{(2)}_0=0$, that is:

\n (1) $B^1_{\;1}(0)=B^2_{\;1}(0)=0$ (no relation between $\xi^{(1)}_0$ and
$\xi^{(2)}_0$),

\n (2) $\xi^{(2)}_0=k_0\,\xi^{(1)}_0$ (with $k_0$ a bosonic constant).

\n The leading singularity is fixed by introducing the values found in the
bosonic-core analysis (corresponding to the `body piece', i.e., without the
nilpotent part, of the bosonic components) and verify the possible solutions
for $r$.

Before pursuing, the exact meaning of this computation should be clarified.
The goal is to fix the leading singularity of the fermionic field for those
5 particular cases for which the bosonic-core analysis manifestly shows the
Painlev\'e
property. We thus look for the solutions (type-(1) or (2)) of (\ferru) for
the special values of the parameters $c,\alpha,\,\beta,\, u_0$ and $w_0$
given in section 4, appropriate to each possibility. The solutions with
negative resonances will not be considered. Now, let us eliminate a possible
source of ambiguity in our procedure: {\it a
priori}, the values of $u_0$ and $w_0$ entering in (\ferru)
should be those pertaining to the complete system, incorporating
the fermions. However, as mentioned above, only the non-nilpotent
parts are considered. The reason for this is that since the nilpotent piece
can be eliminated by an appropriate multiplication, the bosonic core must
also satisfy (\sery).

The solutions to case (1) are:

\+ (I) & $r=-2$, \cr

\+ (II) & $r=0,-2$, \cr

\+ (III) & $r=-2$, \cr

\+ (IV) & $r=0$, \cr

\+ (V) & $r=2,-2,-3$. \cr

For case (2), equations (\ferru) for $n=0$ can be written
$$
B^1_{\;1}(0)-k_0 B^2_{\;1}(0)=0,\qquad \quad
k_0^2 B^1_{\;1}(0) +k_0 B^2_{\;1}(0)=0\, .\eq$$
The compatibility of these equations forces $k_0^2=-1$ or $k_0=\pm i$. The
constant $k=\pm i$ that appear in the expression of the bosonic component
$w_0$ (cf. section 4) can thus be either $k=\pm k_0$; both cases need thus
to be considered (the precise relation being fixed by the resonance
equations). We then find the following possible solutions for $r$:

\+ (I) & $k=+k_0$: && $r=2,-2,-3$, \cr
\+ & $k=-k_0$: && $r=0,-1,-2$, \cr

\+ (II) & $k=+k_0$: && $r=0,-2,-4$, \cr
\+ & $k=-k_0$: && $r=2,0,-2$, \cr

\+ (III) & $k=+k_0$: && $r=-1,-2,-3$, \cr
\+ & $k=-k_0$: && $r=2,0,-2$, \cr

\+ (IV) & $k=+k_0$: && $r=0,-1,-2$, \cr
\+ & $k=-k_0$: && $r=2,0,-5$, \cr

\+ (V) & $r=2,-2,-3$. \cr

\n Observe that the type-(1) solutions for $r$ are recovered as the
intersection of the two set of solutions in each case: this is clear since
in case (1) we do not assume any special relation between $k$ and $k_0$; it
should then hold for all possibilities, in particular when $k=k_0$ and
$-k_0$. In the following, we can thus restrict ourself to type-(2)
solutions.

In order to uniquely fix the value of $r$ (and, thereby, the value of $k$
appropriate to each case), we must consider the resonance equations. Since
the bosonic resonances are solutions of $A(n)=\det|A^i_{\;j}(n)|=0$, the
fermionic resonances are necessarily given by
$$B(n)=\det|B^i_{\;j}(n)|=0 \qquad (i,j=1,2).\eq$$
Inserting the values already found for $r$, $u_0$ and $w_0$, the roots of
$B(n)$ should then lead to the resonance levels for the fermionic fields.
The idea is to select $r$ by requiring the corresponding polynomial $B(n)$
to have only
integer roots. The explicit form of these
polynomials is

\+ (I) & $r=2$: & $B(n)=n(n-2)(n-3)(n-4)^2(n-5)$ \cr
\+ & $r=0$: & $B(n)=(n^5-4n^4-n^3+16n^2-12n+36)(n-2)$ \cr
\+ & $r=-1$: & $B(n)=(n^5+n^4-7n^3-n^2+6n+54)(n-1)$ \cr
\+ & $r=-2$: & $B(n)=(n^5+6n^4+7n^3-6n^2-8n+72)n$ \cr
\+ & $r=-3$: & $B(n)=(n^5+11n^4+41n^3+61n^2+30n+90)(n+1)$ \cr

\+ (II) & $r=2$: & $B(n)=n(n-2)^2(n-4)^2(n-6)$ \cr
\+ & $r=0$: & $B(n)=(n^4-4n^3-4n^2+52n-54)n(n-2)$ \cr
\+ & $r=-2$: & $B(n)=(n^4+4n^3-4n^2+56n-36)(n+2)n$ \cr
\+ & $r=-4$: & $B(n)=(n^4+12n^3+44n^2+156n+54)(n+4)(n+2)$ \cr

\+ (III) & $r=2$: & $B(n)=n(n-2)(n-3)(n-4)^2(n-5)$ \cr
\+ & $r=0$: & $B(n)=(n^4-2n^3-5n^2+42n-18)(n-2)^2$ \cr
\+ & $r=-1$: & $B(n)=(n^5+n^4-7n^3+53n^2-48n-{81 \over 2})(n-1)$ \cr
\+ & $r=-2$: & $B(n)=(n^4+7n^3+14n^2+80n+108)n(n-1)$ \cr
\+ & $r=-3$: & $B(n)=(n^5+11n^4+41n^3+151n^2+210n-{225 \over 2})(n+1)$ \cr

\+ (IV) & $r=2$: & $B(n)=n(n-2)^2(n-3)(n-4)(n-7)$ \cr
\+ & $r=0$: & $B(n)=(n+2)n^2(n-1)(n-2)(n-5)$ \cr
\+ & $r=-1$: & $B(n)=(n+3)(n+1)^2(n-1)n(n-4)$ \cr
\+ & $r=-2$: & $B(n)=(n+4)(n+2)^2(n+1)n(n-3)$ \cr
\+ & $r=-5$: & $B(n)=(n+7)(n+5)^2(n+4)(n+3)n$ \cr

\+ (V) & $r=2$: & $B(n)=n^2(n-4)^2(n-5)^2$ \cr
\+ & $r=-2$: & $B(n)=(n+4)^2n^2(n-1)^2$ \cr
\+ & $r=-3$: & $B(n)=(n+5)^2(n+1)^2n^2$ \cr

\n We can already eliminate all cases for which there are non-integer roots.
This leaves us with $r=2$ as the only possibility for cases (I), (II) and (III)
while we have some other possibilities for cases (IV) and (V). 
However, we argue in section 5 that for the other possibilities we can restrict to
$r=2$ 
(moreover, this amounts to  restrict the study
to the principal families).

The situation concerning the leading fermionic singularity is thus somewhat
peculiar: we essentially keep track of all possibilities and determine the
particular values which ensure integer-valued resonances. Quite
interestingly, the same value for $r$ is singled out in all
cases when we restrict to principal family solutions. Actually, this value
corresponds precisely to the one that
follows from a naive consideration where the fermionic terms, in
the bosonic evolution equations, have a dominant singular
behavior comparable to that of the leading bosonic terms.

\vskip0.3cm
\centerline{\bf ACKNOWLEDGEMENTS}
The work of S.B. was supported by NSERC (Canada), through an
Undergraduate Research Student Award and that of P.M. by
NSERC (Canada) and FCAR (Qu\'ebec).

\vskip0.3cm
\centerline{\bf REFERENCES}
\immediate\closeout\refs \vskip 0.5cm
\message{References}\input references
\vfill\eject
\end